\documentclass[11pt]{article} 
\usepackage{amsfonts,amssymb,latexsym}
\usepackage{graphicx,graphics,amssymb,epsf,rotate} 
\usepackage{axodraw,cite,mathrsfs}
\evensidemargin=\oddsidemargin 
\def\Eqn#1{Eq.\ (\ref{#1})}
\def\Eqs#1#2{Eqs.\ (\ref{#1}) and (\ref{#2})}
\def\3Eqs#1#2#3{Eqs.\ (\ref{#1}), (\ref{#2}) and (\ref{#3})}
\def\fig#1{Fig.~\ref{#1}}
\def\ms{\mathscr}

\def\mchi{M_{\tilde{\chi}}}

\def\lsim{\mathrel{\raise-0.8ex\hbox{$\stackrel{\textstyle <}\sim$}}}
\def\gsim{\mathrel{\raise-0.8ex\hbox{$\stackrel{\textstyle >}\sim$}}}

\begin{document}

\begin{center}
{\Large \bf A reappraisal of spontaneous $R$-parity violation} \\[1cm]
{\sf Gautam Bhattacharyya} and {\sf Palash B. Pal}  \\[10pt]
{\small {\em Saha Institute of Nuclear Physics, 1/AF Bidhan
    Nagar, Kolkata 700064, India }}
\normalsize
\end{center}

\begin{abstract}
  In this short reappraisal of spontaneous lepton number violation in
  a supersymmetric scenario implemented through singlet sneutrino
  vacuum expectation value (VEV), we contribute with two new things in
  the context where the lepton number symmetry is global: (i) provide
  explicit expressions of $R$-parity violating couplings in terms of
  the neutrino Yukawa couplings and the singlet sneutrino VEV, and
  (ii) estimate the limit on this VEV using the current knowledge of
  the light neutrino mass and the astrophysical constraint on the
  Majoron-electron coupling.  Besides, we put updated constraints on
  the VEV and Yukawa coupling of the singlet superfield when lepton
  number is gauged.

\vskip 5pt \noindent
\texttt{Key Words: $R$-parity, Supersymmetry, Majoron}
\end{abstract}

\section{Introduction}
It is well-known that if the supersymmetric partners of all standard model
(SM) particles are introduced in a theory and one constructs the most general
Lagrangian that is invariant under supersymmetry and the SM gauge symmetry,
the Lagrangian contains terms which violate both lepton number ($\mathcal{L}$)
and baryon number ($\mathcal{B}$). The {\em explicit} $\mathcal{L}$- and
$\mathcal{B}$-violating parts appear in the superpotential as:
\begin{eqnarray}
\ms W \supset \sum_{abc} \lambda_{abc} L_a L_b \hat E_c +
\sum_{abc} \lambda'_{abc} L_a Q_b \hat D_c +
\sum_{abc} \lambda''_{abc} \hat U_a \hat D_b \hat D_c + 
\sum_a \mu_a L_a H_u \,.
\label{rpvterms}
\end{eqnarray}
Above, all superfields are left-chiral, and the subscripts $a,b,c$ are
generation indices on lepton doublet fields $L$, quark doublet fields $Q$, and
SU(2)-singlet charged fields $E$, $U$ and $D$, in obvious notations.  The hat
on a superfield means that the left-chiral fermion part of that superfield is
the antiparticle of fermion whose name is suggested in the letter denoting the
superfield.  These terms also violate $R$-parity, defined by
$(-1)^{3\mathcal{B} + \mathcal{L} + 2 \mathcal{S}}$, where $\mathcal{S}$ is
the spin of the particle. The antisymmetry in the first 
two generation
indices of $\lambda$ and in the last two indices of 
$\lambda''$ suggests $\lambda_{abc} = - \lambda_{bac}$
and $\lambda''_{abc} = - \lambda''_{acb}$. Clearly, there are 9
$\lambda$-type, 27 $\lambda'$-type and 9 $\lambda''$-type trilinear, plus 3
$\mu$-type bilinear, $R$-parity violating (RPV) couplings. These 48 new
couplings add further twists and complications to phenomenology and
bring in more uncertainty to theoretical predictions.

Our aim in this paper is to explore a restrictive scenario in which there will
be much fewer RPV couplings, thus offering more predictivity.  For
definiteness, we assume that these couplings are generated by {\em
  spontaneous} $\mathcal{L}$ violation. This immediately rules out the
$\mathcal{B}$-violating $\lambda''$-type couplings. Now, we recall that Aulakh
and Mohapatra \cite{Aulakh:1982yn} were the first to have implemented the idea
of spontaneous violation of $\mathcal{L} = 1$ global lepton number in a
supersymmetric context through the VEV of the sneutrino component of a lepton
doublet superfield.  A testable feature of this realisation was a photino
mediated contribution to neutrinoless double beta decay. Neutrinos were
predicted to be massless at tree level, with a suggestion that very small
masses ($m_{\nu} \sim (10^{-5} - 10^{-8})$ eV) are induced at one-loop order
through a combination of supersymmetry breaking and lepton number violation. A
follow-up study \cite{Nieves:1983kn} revealed that if supersymmetry breaking
terms include trilinear scalar couplings and gaugino Majorana masses, neutrino
mass would be generated at the tree level itself, with a special property that
even with three generations only one non-vanishing mass eigenvalue would
emerge at tree level.  The mechanism of \cite{Aulakh:1982yn} would indeed
induce two other small masses at one-loop order. Other implications of this
scenario were studied in the context of matter-enhanced solar neutrino
oscillation \cite{Santamaria:1988zm, Santamaria:1987uq, Santamaria:1988ic}.

An important feature of spontaneous $\mathcal{L}$ violation is the existence
of Majoron ($J$), which is a physical massless Nambu-Goldstone boson arising
from the imaginary part of the sneutrino field that acquires a VEV.  The mass
of the real scalar ($\rho$), associated with the Majoron, in the doublet
Majoron scenario turns out to be very small leading to unacceptably large $Z
\to \rho J$ decay, which is ruled out by the LEP data on $Z$ boson decay
width. In fact, gauge non-singlet Majoron models are all strongly disfavored
by electroweak precision measurements \cite{GonzalezGarcia:1989zh,
  Romao:1989yh}. Subsequently, {\em singlet} Majoron scenarios were proposed
in the supersymmetric context.  In some of these models, lepton number was
spontaneously broken by the VEV of a field carrying one unit of lepton number
\cite{Masiero:1990uj, Romao:1991ex, Romao:1992vu}, and in some others by a
field carrying two units of the lepton number \cite{Giudice:1992jg}, like the
non-supersymmetric models of spontaneous lepton number violation
\cite{Chikashige:1980qk, Chikashige:1980ui}.  Since the first kind,
i.e. $\Delta \mathcal{L} = 1$ violation, is a speciality of supersymmetric
models that non-supersymmetric models do not have, we take it up for our work
here.

In this context, we present explicit expressions of $\lambda_{abc}$,
$\lambda'_{abc}$ and $\mu_a$ couplings in terms of the Yukawa couplings of the
general superpotential and the singlet neutrino VEV. Furthermore, we provide
new bounds on the sneutrino VEV and the generic neutrino Yukawa couplings from
astrophysical considerations of stellar energy loss.

\section{$R$-parity violating couplings}
Our model has, apart from the superfields in the MSSM, gauge singlet
superfield $\hat N_a$, one for each generation.  These fields carry lepton
number $\mathcal{L} = -1$, and hence their VEVs ($V_a$) would spontaneously
break lepton number.  The superpotential of this model can be written as
\begin{eqnarray}
\ms W &=& \sum_{ab} h^{(u)}_{ab} Q_a \hat U_b H_u +
\sum_{ab} h^{(d)}_{ab} Q_a \hat D_b H_d 
 +
\sum_{ab} h^{(l)}_{ab} L_a \hat E_b H_d +
\sum_{ab} h^{(N)}_{ab} L_a \hat N_b H_u + \mu H_u H_d \,.
\label{superpot}
\end{eqnarray}
We assume that the VEVs of $\hat N_a$ are generated by the same
mechanism as in \cite{Masiero:1990uj, Romao:1991ex, Romao:1992vu}
which require the existence of two more gauge singlet superfields
(with $\mathcal{L} = +1$ and $0$, respectively). Though we implicitly
acknowledge their existence we do not explicitly display how those two
additional singlets appear in the superpotential, whose {\em raison
  d'\^{e}tre} is to provide the $V_a$'s.  Beside that,
Eq.~(\ref{superpot}) is the most general gauge invariant
superpotential that also conserves lepton number before the scalar
components of $\hat N_a$ go to the vacuum. It is not difficult to
realize that the VEVs $V_a$ by themselves cannot break
supersymmetry. Also note that apart from the term containing the $\hat
N$ fields, the superpotential corresponds exactly to that of
$R$-parity conserving minimal supersymmetry.

\begin{figure}
\begin{center}

\begin{picture}(120,100)(-60,-50) 
\SetWidth{3}
\ArrowLine(-60,-40)(-30,0)
\Text(-60,-50)[]{$L$}
\ArrowLine(-60,40)(-30,0)
\Text(-60,50)[]{$\hat N$}

\ArrowLine(-4,0)(-30,0)
\Text(-17,6)[b]{$H_u$}
\ArrowLine(4,0)(30,0)
\Text(17,6)[b]{$H_d$}
\Vertex(0,0)4

\ArrowLine(60,-40)(30,0)
\Text(60,-50)[]{$L$}
\ArrowLine(60,40)(30,0)
\Text(60,50)[]{$\hat E$}

\Text(0,-50)[]{\large (a)}
\end{picture}
\hspace{2cm}
\begin{picture}(120,100)(-60,-50) 
\SetWidth{3}
\ArrowLine(-60,-40)(-30,0)
\Text(-60,-50)[]{$L$}
\ArrowLine(-60,40)(-30,0)
\Text(-60,50)[]{$\hat N$}

\ArrowLine(-4,0)(-30,0)
\Text(-17,6)[b]{$H_u$}
\ArrowLine(4,0)(30,0)
\Text(17,6)[b]{$H_d$}
\Vertex(0,0)4

\ArrowLine(60,-40)(30,0)
\Text(60,-50)[]{$Q$}
\ArrowLine(60,40)(30,0)
\Text(60,50)[]{$\hat D$}

\Text(0,-50)[]{\large (b)}
\end{picture}

\end{center}

\caption{\small\em Generation of $LL\hat E$ and $LQ\hat D$ operators
  when the scalar part of $\hat N$ acquires a VEV. The thick lines
  denote superfields.  Generation indices have been suppressed.  The
  blob in the middle of the internal line implies that it involves the
  $\mu$-term of the superpotential.}
\label{f:superdiags}
\end{figure}
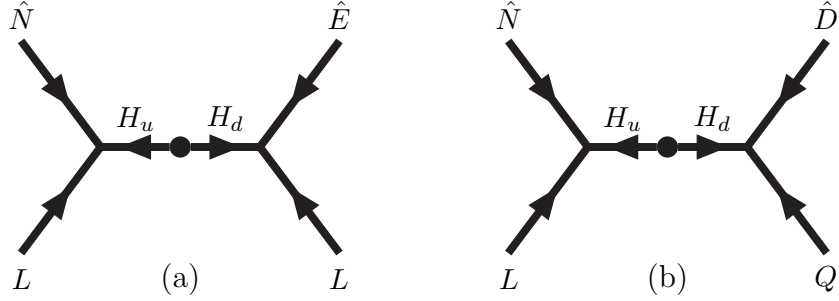
The $\mathcal{L}$-violating operators are generated as soon as the
VEVs $V_a$ are induced.  \fig{f:superdiags}a will generate the
$\lambda$ terms, whereas \fig{f:superdiags}b will generate $\lambda'$
terms of Eq.~(\ref{rpvterms}).  The important point is that, these
couplings will now be determined by Yukawa couplings and the sneutrino
VEVs.  It should be noted that when $\hat N$ acquires a VEV, the
internal line in \fig{f:superdiags} is necessarily Higgsino.  If we
assume that the Higgsino mass parameter $\mu$ is a few hundred GeV or
more, one can effectively write a contact interaction from
\fig{f:superdiags}.  When the singlet sneutrinos acquire VEVs, we
obtain
\begin{eqnarray}
\lambda_{abc} &\simeq& \sum_d {V_d \over
  \mu} \; \left( h^{(N)}_{ad} h^{(l)}_{bc} -  h^{(N)}_{bd} h^{(l)}_{ac} \right)
\,,   
\label{lambda}\\ 
\lambda'_{abc} &\simeq& \sum_d {V_d \over \mu} \; 
h^{(N)}_{ad} h^{(d)}_{bc} \,.
\label{lambda'}
\end{eqnarray}
The two terms in case of $\lambda_{abc}$ arise depending on whether
$L_a$ or $L_b$ appears in the same vertex with the $\hat N_d$
superfield in \fig{f:superdiags}a.  The bilinear lepton number
violating terms, shown in \Eqn{rpvterms}, also arise in this model
from the $h^{(N)}$ terms of the superpotential in \Eqn{superpot} when
the scalar component of $\hat N$ acquires a VEV:
\begin{eqnarray}
\mu_a = \sum_b h^{(N)}_{ab} V_b \,.
\label{mua}
\end{eqnarray}

The origin of the relative minus sign between the two terms in \Eqn{lambda}
can be understood by keeping the SU(2) indices.  If we denote the SU(2) index
carried by $L_a$ and $L_b$ by $\alpha$ and $\beta$ respectively, and put the
SU(2) indices $\gamma$ and $\delta$ on the internal $H_u$ and $H_d$ superfield
lines in \fig{f:superdiags}, then the diagram with $L_a$ coupling directly to
$\hat N_d$ will contain the SU(2) factor
\begin{eqnarray}
\varepsilon_{\alpha\gamma} \varepsilon_{\gamma\delta}
\varepsilon_{\beta\delta} =  \varepsilon_{\beta\alpha} \,,
\end{eqnarray}
whereas the other diagram, obtained by interchanging $L_a$ and $L_b$,
should contain
\begin{eqnarray}
\varepsilon_{\beta\gamma} \varepsilon_{\gamma\delta}
\varepsilon_{\alpha\delta} = \varepsilon_{\alpha\beta} \,.
\end{eqnarray}
Hence the minus sign in \Eqn{lambda}, which makes the coupling antisymmetric
in the indices $a,b$.

We can now estimate how many unknown parameters are present in the
$\mathcal{L}$-violating sector.  Without any loss of generality, the
couplings $h^{(d)}$ and $h^{(l)}$ can be taken diagonal in the generation
indices, and they can be made real.  In this case, these couplings are
proportional to the masses of the down-type quarks and charged leptons, and
are therefore known.  The couplings $h^{(u)}$ are irrelevant for our
discussion since they do not appear in the expressions of
\3Eqs{lambda}{lambda'}{mua}.  These will contain the up-type quark masses and
the parameters of the Cabibbo-Kobayashi-Maskawa matrix.  For our purpose, the
relevant unknown parameters appear from $h^{(N)}$, and they are 9 in number.
Besides, there are the three VEVs $V_a$.  To be more precise, there is
actually only one independent VEV of the singlet sneutrino fields, since we
can always make suitable linear combinations of the three $\hat N$ fields
leading to the occurrence of a single VEV.  This makes a total count of 10,
instead of the 39 $\mathcal{L}$-violating parameters appearing in
\Eqn{rpvterms}.

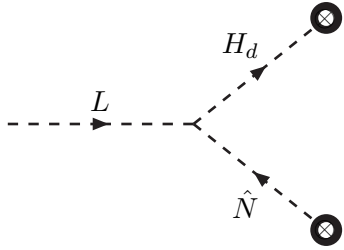
\begin{figure}
\def\VEVblob(#1,#2){\Vertex(#1,#2)6      
{\SetColor{White} \Vertex(#1,#2)3}\Text(#1,#2)[]{$\otimes$}}
  \centering
  \begin{picture}(150,100)(30,-50)
\SetWidth{1}
    \DashArrowLine(30,0)(100,0)4
    \Text(65,5)[b]{$L$}
    \DashArrowLine(100,0)(150,40)4
    \Text(125,25)[br]{$H_d$}
    \VEVblob(150,40)
    \DashArrowLine(150,-40)(100,0)4
    \Text(125,-25)[tr]{$\hat N$}
    \VEVblob(150,-40)
  \end{picture}
  \caption{\small\em Diagrammatic way of seeing how a doublet sneutrino
    gets a VEV.  The dashed lines denote the neutral scalar fields
    contained in the supermultiplets, and the cross-hatched blobs
    denote their VEVs.}
  \label{f:vL}
\end{figure}

\section{Sneutrino VEV and Majoron-electron coupling}
The Majoron-electron coupling arises both at tree level and at one loop order
from different interactions, which could be of similar magnitude. The tree
contribution originates from the fact that a non-vanishing $V$, the generic
VEV of a $\hat N$ field, must accompany a non-vanishing $v_L$, the VEV of a
doublet sneutrino.  This can be seen most easily from the fact that in the
scalar potential of the model, the $F$-term of the $H_u$ field contains a term
of the form $\mu h^{(N)} L \hat N H_d^\dagger$, where in this expression only
the scalar components of each superfield is implied and generation indices are
suppressed.  The diagram in \fig{f:vL} now clearly shows that the magnitude of
$v_L$ should be given by
\begin{eqnarray} 
v_L \approx \frac{\mu h^{(N)} V v_d}{M_S^2} \, ,
\label{vL}
\end{eqnarray}
where $h^{(N)}$ is the generic Yukawa coupling involving the $\hat N$
fields, $v_d$ is the Higgs VEV contained in $H_d$, and $M_S$ is a
generic doublet sneutrino mass.  The above expression can also be
appreciated directly from potential minimization.  In the tadpole
equation $\partial V/\partial v_L=0$, the trilinear term $\mu h^{(N)}
L \hat N H_d^\dagger$ will provide a contribution $\mu h^{(N)} V v_d$
for the left-hand side, while the soft mass term $M_S^2 L^\dagger L$
will yield $M_S^2v_L$.  The minimization condition thus gives \Eqn{vL}.  

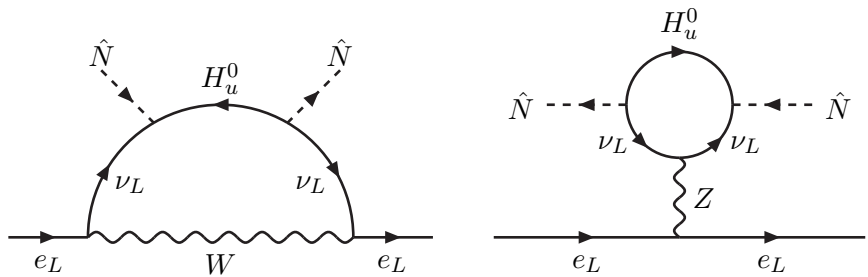
\begin{figure}[b]
\begin{center}
\begin{picture}(180,90)(-90,-10)
\SetWidth{1}

\ArrowLine(-80,0)(-50,0)
\Text(-65,-10)[]{$e_L$}
\ArrowLine(50,0)(80,0)
\Text(65,-10)[]{$e_L$}
\ArrowArcn(0,0)(50,60,0)
\Text(-40,20)[l]{$\nu_L$}
\ArrowArc(0,0)(50,60,120)
\Text(0,60)[]{$H_u^0$}
\ArrowArcn(0,0)(50,180,120)
\Text(40,20)[r]{$\nu_L$}
\Photon(-50,0)(50,0)28
\Text(0,-10)[]{$W$}
\DashArrowLine(-45,63)(-25,43)3
\Text(-45,70)[]{$\hat N$}
\DashArrowLine(25,43)(45,63)3
\Text(45,70)[]{$\hat N$}

\end{picture}
\begin{picture}(160,100)(-80,-10)
\SetWidth{1}

\ArrowLine(-70,0)(0,0)
\Text(-35,-10)[]{$e_L$}
\ArrowLine(0,0)(70,0)
\Text(35,-10)[]{$e_L$}
\Photon(0,0)(0,30)23
\Text(5,15)[l]{$Z$}
\ArrowArcn(0,50)(20,180,0)
\Text(0,75)[b]{$H_u^0$}
\ArrowArc(0,50)(20,180,270)
\Text(-25,35)[]{$\nu_L$}
\ArrowArc(0,50)(20,270,360)
\Text(25,35)[]{$\nu_L$}
\DashArrowLine(-20,50)(-50,50)3
\Text(-55,50)[r]{$\hat N$}
\DashArrowLine(50,50)(20,50)3
\Text(55,50)[l]{$\hat N$}

\end{picture}
\end{center}
\caption{\small\em Effective operators that give rise to a coupling
  between the electron and the Majoron when one of the external scalar
  lines goes to the vacuum.}\label{f:eJ}
\end{figure}

The non-zero value of the doublet sneutrino VEV induces, from the
supergraph shown in \fig{f:superdiags}a, a tree-level electron-Majoron
coupling.  Assuming, for the sake of illustration, that the soft mass
of $H_d$ is of the same order as $\mu$, this coupling is given by
\begin{eqnarray} 
g_{eeJ}^{\rm tree} \approx \frac{1}{M_S^2} \Big( h^{(N)} \Big)^2 V m_e \, ,
\label{eeJ-tree}
\end{eqnarray}
where the factor $m_e$, equal to $h^{(l)}$ times $v_d$, ensures a
chirality-flipping coupling.  The loop induced contribution to the
electron-Majoron coupling arises from the diagrams shown in
\fig{f:eJ}, when one of the external $\hat N$-lines obtains a VEV.  A
rough estimate of the coupling thus generated is of the order
%
\begin{eqnarray}
  g_{eeJ}^{\rm loop} \approx \frac{g^2}{16 \pi^2 M_0^2} 
\Big( h^{(N)} \Big)^2 V m_e \,,
\label{eeJ-loop}
\end{eqnarray}
where $M_0$ is the heaviest mass in the diagram, either of the $Z$
boson or of the neutralino (through its Higgsino component).  The
magnitude of the tree and loop contributions to $g_{eeJ}$ could be of
the same order, or one may dominate over the other, depending on the
magnitude of the parameters $M_S$ and $M_0$.  A cancellation between
the two contributions is unlikely and we avoid any such fine-tuning.

\section{Combined astrophysical and neutrino mass constraints}
There are stringent astrophysical constraints on any Majoron model.
Majoron emission leads to stellar energy loss, and singlet Majorons
may be emitted via Compton-like processes $\gamma + e \to e + J$. The
allowed leaking of stellar energy can be translated into a bound on
the singlet sneutrino VEV.  In fact, it turns out that Majoron
coupling to the electron should be less than about $10^{-10}$
\cite{Dicus:1978fp, Fukugita:1982gn, Pantziris:1986dc, Chanda:1987ax}
from a study of the main sequence stars.  Putting $M_S \sim 100$\,GeV
in the tree level contribution \Eqn{eeJ-tree}, the Majoron emission
bound implies
\begin{eqnarray}
\Big( h^{(N)} \Big)^2 \, V \lsim 2 \; \mbox{MeV} \, ,  
\label{astro-bound-tree}
\end{eqnarray}
while from the loop contribution \Eqn{eeJ-loop}, for $M_0 = 100$\,GeV,
the bound turns out to be
\begin{eqnarray}
\Big( h^{(N)} \Big)^2 \, V \lsim 1 \; \mbox{GeV} \,.
\label{astro-bound-loop}
\end{eqnarray}
Although the bounds in \Eqs{astro-bound-tree}{astro-bound-loop} are
quite different, both are independently significant, as the scalar
mass ($M_S$) involved in \Eqn{eeJ-tree} can be much larger than the
neutralino (or, the $Z$ boson) mass ($M_0$) in \Eqn{eeJ-loop}.  The
above bounds can be more stringent if, instead of main sequence stars,
we use red giant stars, which give $g_{eeJ} \lsim 3\cdot 10^{-13}$
\cite{Raffelt:1990yz}.  However, we use the constraints from main
sequence stars which are more reliably understood.

It is interesting to note from \Eqs{eeJ-tree}{eeJ-loop} that the
electron coupling to the Majoron is {\em directly} proportional to the
lepton number breaking VEV.  In contrast, the same coupling is {\em
  inversely} proportional to the lepton number breaking VEV in the
non-supersymmetric singlet Majoron model \cite{Chikashige:1980qk,
  Chikashige:1980ui}.  The reason for the difference is that in the
non-supersymmetric case, where lepton number symmetry is broken by the
VEV of a scalar field carrying two units of lepton number, heavy
singlet neutrinos whose mass is of the same order as their VEVs float
in the loop causing propagator suppression.  In our case, $V$ appears
only in the numerator when an $\hat N$ is replaced by its VEV.

\begin{figure}
\begin{center}
  \begin{picture}(250,80)(0,-10)
    \SetWidth{1}
\Line(0,0)(250,0)
\ArrowLine(0,0)(50,0)
\DashArrowLine(50,50)(50,0)3
\ArrowLine(100,0)(50,0)
\DashArrowLine(100,50)(100,0)3
\Photon(100,0)(150,0){-2}{4.5}
\DashArrowLine(150,50)(150,0)3
\ArrowLine(150,0)(200,0)
\DashArrowLine(200,50)(200,0)3
\ArrowLine(250,0)(200,0)
\Text(25,-10)[c]{$L_a$}
\Text(75,-10)[c]{$H_u$}
\Text(125,-10)[c]{$\tilde Z_{}$}
\Text(175,-10)[c]{$H_u$}
\Text(225,-10)[c]{$L_b$}
\Text(50,55)[b]{$\hat N$}
\Text(100,55)[b]{$H_u$}
\Text(150,55)[b]{$H_u$}
\Text(200,55)[b]{$\hat N$}
  \end{picture}
\end{center}

\caption{\small\em Schematic diagram showing how a neutrino acquires
    a tree level mass in our model after the $\hat N$ fields and the
    neutral $H_u$ acquires VEVs.}\label{f:tree-mass}
\end{figure}
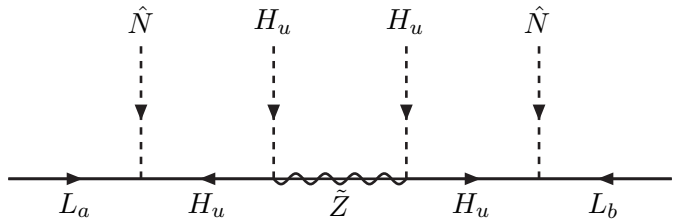

We now discuss how neutrino mass is generated in our scenario.
\fig{f:tree-mass} is a diagrammatic representation of $\Delta {\cal L}
= 2$ Majorana mass generation at tree level.  It gives
\begin{eqnarray}
m_{ab} \propto h_{ac}^{(N)} h_{bd}^{(N)} V_c V_d \, .
\label{rankone}
\end{eqnarray} 
Since the mass matrix is of rank one, only one non-vanishing eigenvalue will
emerge.  
This is not surprising, as one can always perform a basis rotation in
flavor space to put the VEV only along one direction.  In fact, what
we discussed is nothing but the neutrino mass generation through
bilinear RPV couplings.  Accurate expressions of the tree level
neutrino mass induced by bilinear RPV couplings can be found, for
example, in \cite{Romao:1991ex}.  For our purpose, it is enough to use
the approximate expression of neutrino mass suggested by
\fig{f:tree-mass}:
\begin{eqnarray} 
  m_\nu \sim g^2 \left(h^{(N)}\right)^2 \frac{v_u^2 V^2}
    {\mchi^3} \,,  
\label{mnu}
\end{eqnarray} 
where $v_u$ denotes the VEV of the scalar component of $H_u$, and $\mchi$ is a
generic neutralino mass capturing the effects of the Zino and Higgsino
propagators in \fig{f:tree-mass}, where it is implicitly assumed that $\mchi
\sim \mu$.  Even though we do not yet precisely know the absolute magnitude of
neutrino mass, we make a reasonable guess by putting $m_\nu = 0.1$\,eV in
\Eqn{mnu}. We further assume that $\mchi \sim v_u \sim 100$\,GeV, and obtain
\begin{eqnarray}
h^{(N)} V \sim 2 \cdot 10^{-4} ~{\rm GeV} \, . 
\label{numass_constraint}
\end{eqnarray} 
If we compare \Eqn{numass_constraint} with the astrophysical bound in
\Eqn{astro-bound-tree} obtained from the tree level contribution to
electron-Majoron coupling, assuming that the neutrino Yukawa couplings
involved with Majoron emission and neutrino mass generation are of the same
order, we obtain two limits:
\begin{eqnarray} 
V \gsim 20\,{\rm keV} \,, \qquad h^{(N)} \lsim 10 \, .
\label{V-h-bounds-tree}
\end{eqnarray} 
On the other hand, if we compare \Eqn{numass_constraint} with 
\Eqn{astro-bound-loop}, the bound on the electron-Majoron coupling from the
loop process, we obtain 
\begin{eqnarray} 
V \gsim 40 \,{\rm eV} \,, \qquad h^{(N)} \lsim 5 \cdot 10^3 \, .
\label{V-h-bounds-loop}
\end{eqnarray} 

The upper bound on $h^{(N)}$ is not particularly useful if the theory has to
be perturbative.  However, it is interesting to observe that while the
astrophysical bounds in \Eqs{astro-bound-tree}{astro-bound-loop} are {\em
  upper} bounds on a combination of the neutrino Yukawa coupling and the
$\mathcal{L}$-violating VEV, finally we obtain {\em lower} bounds on the
latter --- see \Eqs{V-h-bounds-tree}{V-h-bounds-loop} --- if we assume some
reasonable value of the light neutrino mass.  This is because
\3Eqs{eeJ-tree}{eeJ-loop}{mnu} imply an order-of-magnitude relation between
Majoron coupling and neutrino mass:
\begin{eqnarray}
g_{eeJ}^{\rm tree} \sim \; {\mchi^3 \over g^2 v_u^2 M_S^2} 
\; {m_e m_\nu \over V}
\,,
\qquad 
g_{eeJ}^{\rm loop} \sim {1 \over 16\pi^2} \; {\mchi^3 \over
  v_u^2 M_0^2} \; {m_e m_\nu \over V} \,,
\end{eqnarray}
where, we recall from \Eqn{eeJ-loop} that $M_0$ is either $\mchi$ or
$M_{Z}$, whichever is larger.  Such a proportionality between neutrino
mass and Majoron-electron coupling occurs also in other singlet
Majoron models \cite{MohaPal3ed}.

It should be noted that the electron-Majoron coupling in
\Eqs{eeJ-tree}{eeJ-loop} contains $\big( h^{(N)} \big)^2$.  In writing
this, we have suppressed the generation indices.  More explicitly, the
combination that actually appears is $\sum_i h^{(N)}_{ei}
h^{(N)}_{ei}$.  The same combination is constrained from the
electron-neutrino mass, whose expression appears in \Eqn{mnu}.  This
implies further constraints on other combinations, involving different
charged leptons, through neutrino mixing
parameters~\cite{Hirsch:2009ee}.

Now we turn our attention to the $\lambda$ and $\lambda'$ couplings in
\Eqs{lambda}{lambda'} and see what information on $h^{(N)}$ we get from
them. Using the neutrino mass constraint in \Eqn{numass_constraint}, the
dimensionless prefactor $h^{(N)}V/\mu \sim 10^{-6}$, for $\mu \sim 100$ GeV,
provides sufficient suppression to $\lambda$ and $\lambda'$ couplings, in
addition to those coming from charged lepton (or, down-type quarks) Yukawa
couplings, to meet all experimental constraints \cite{Bhattacharyya:1996nj,
  Bhattacharyya:1997vv, Dreiner:1997uz, Chemtob:2004xr, Barbier:2004ez}.  As a
result, we can keep the $h^{(N)}$ matrix elements to be all order unity.  We
note that, unlike the trilinear $\lambda$ or $\lambda'$ couplings, the
bilinear $\mu_a$ parameters do not pick up the extra suppression from charged
lepton Yukawa couplings, and we may expect that the corresponding bilinear
soft terms are not suppressed either.  This observation helps us to face an
important question at this stage: how do we produce an acceptably large {\em
  second} neutrino mass eigenvalue? This could be induced by Grossman-Haber
loops \cite{Grossman:1997is} which contribute to neutrino mass through the
$\mathcal{L}$-violating bilinear soft terms. In these loops there are gauge
couplings at the neutrino vertices, and there are two types of
$\mathcal{L}$-violating interactions (e.g. slepton-Higgs or
neutrino-neutralino mixing) inside the loop giving rise to $\Delta \mathcal{L}
= 2$ interactions. Addition of these loops to \Eqs{rankone}{mnu} breaks the
rank one structure of the mass matrix, but one eigenvalue still remains
zero. This is very much consistent with the neutrino oscillation data, which
do not any way compel us to consider a non-vanishing {\em third} mass
eigenvalue. The generation of the latter requires the relevant $\lambda$ or
$\lambda'$ couplings to be $\sim (10^{-3}-10^{-4})$ for superparticle masses
of order 100 GeV, but in our scenario these couplings are significantly more
suppressed (For a detailed discussion of how RPV couplings generate neutrino
masses and mixing, see, for example, \cite{Abada:2002ju, Abada:2001zh,
  Davidson:2000ne, Davidson:2000uc, Bhattacharyya:1999tv, Rakshit:1998kd,
  Hirsch:2000ef}).

\section{Scenario with gauged lepton number}
\Eqn{superpot} can also be interpreted as the superpotential of a
model where the lepton number symmetry is gauged.  Of course, lepton
number is anomalous, but the combination $\mathcal{B-L}$ is not.  So,
as a simplest example, we can think of a model where the gauge
symmetry consists, apart from the standard model gauge group, of
another factor of $\rm U(1)_{\mathcal{B-L}}$.  This is the same as the
model presented in Ref.~\cite{GonzalezGarcia:1990qf}, where a
different combination of the weak hypercharge and $\mathcal{B-L}$ had
been used to denote this extra symmetry.

Without any loss of generality, we assume that only one singlet
sneutrino acquires VEV, and to avoid confusion with the global
symmetry case, we denote this VEV by $v_R$.  There will be no Majoron
in this case: the Goldstone boson will be eaten up by the extra
neutral gauge boson that is present in this model.  The strength of
$R$-parity violation will be related to the strength of this new gauge
force.  It has been shown \cite{GonzalezGarcia:1990qf} in the context
of an $E_6$-inspired model that
\begin{eqnarray}
M_{Z'}^2 = \frac43 \tan^2\theta_W M_W^2 + 
{25 \over 12} g'^2  v_R^2 \,.
\end{eqnarray}
Using the current experimental lower limit $M_{Z'}>900$\,GeV
\cite{pdg} from the Tevatron $p\bar p$-collider, we obtain the limit
\begin{eqnarray}
v_R > 1.7 \, {\rm TeV} \,.
\label{vRbound}
\end{eqnarray}

Although we cannot use the astrophysical bounds for this model since
there is no Majoron, the neutrino mass formula given in \Eqn{mnu}
still holds, where $V$ should be read as $v_R$.  Using $m_\nu =
0.1$\,eV as before and putting the lower bound on $v_R$ from
\Eqn{vRbound}, we obtain 
\begin{eqnarray}
\mchi \sim \left(h^{(N)}\right)^{2/3} \times 4500\, {\rm TeV} \,.
\end{eqnarray}
Unlike in the previous example with global lepton number symmetry, the
elements of the Yukawa matrix $h^{(N)}$ will now have to be very small
in order to keep the neutralino mass $\mchi$ in the phenomenologically
interesting range of a few hundred GeV to a TeV.

\section{Conclusion}
We have done a few new things in this paper.  In spite of the existence
of a vast literature on spontaneous $R$-parity violation, explicit
expressions of lepton number violating couplings in terms of the
sneutrino VEV and neutrino Yukawa couplings were somehow
missing. Also, only bilinear RPV terms were discussed in the context
of Majoron models so far. As we have shown, trilinear RPV terms would
be generated too. We displayed them in \3Eqs{lambda}{lambda'}{mua}.
These might be particularly useful if one attempts to construct some
flavor models relating different entries of the $h^{(N)}$ matrix,
which can comfortably be of order unity.  Note that using 10
parameters we can predict 39 $R$-parity violating couplings.  The
other new thing that we presented is an explicit estimate of the
bounds on the singlet sneutrino VEV and the generic $h^{(N)}$ by using
the astrophysical constraint on Majoron-electron coupling and the
knowledge of the neutrino mass: see
\Eqs{V-h-bounds-tree}{V-h-bounds-loop}.

We have also noted that when lepton number is gauged, the non-observation of
any additional gauge boson in any collider experiment puts a strong bound on
the singlet sneutrino VEV, which in turn demands that entries of the $h^{(N)}$
matrix have to be small to keep the neutralino masses in the accessible range.

\bigskip

\paragraph* {Acknowledgments:} 
We thank J.\ Romao, especially, for reading the manuscript and making several
important comments and A.\ Vicente for useful discussions.  GB thanks G.~C.\
Branco for inviting him to spend a few weeks at the Instituto Superior
T\'ecnico, Lisbon, where a part of the work was done. GB's visit to Lisbon was
in part supported by Funda\c c\~ ao para a Ci\^encia e a Tecnologia (FCT,
Portugal) through the projects CERN/FP/109305/2009 and CFTP-FCT Unit 777 which
are partially funded through POCTI (FEDER). PBP acknowledges hospitality at
Technische Universit\"at Dortmund during a part of this work. Both authors
acknowledge partial support of the DST-DAAD project INT/DAAD/P-181/2008 and
DAAD-DST PPP grant D/08/04933.


\end{document}